\documentclass[sigconf]{acmart}

\usepackage{booktabs}
\usepackage{latexsym,times}
\usepackage{epsfig}
\usepackage{amsmath}
\usepackage{graphicx,dblfloatfix}
\usepackage{tabularx}
\usepackage{verbatim}
\usepackage{listings}
\usepackage{algorithm}
\usepackage{algpseudocode}
\usepackage{caption}
\usepackage{multirow}
\usepackage{pifont}
\usepackage{balance}

\newenvironment{smallitem}{
   \setlength{\topsep}{0pt}
   \setlength{\partopsep}{0pt}
   \setlength{\parskip}{0pt}
   \begin{itemize}
   \setlength{\leftmargin}{.2in}
   \setlength{\parsep}{0pt}
   \setlength{\parskip}{0pt}
   \setlength{\itemsep}{0pt}}{\end{itemize}}

\begin{document}

\input{epsf}

\pagestyle{plain}

 \title{A Collaborative Filtering Recommender System \\ 
for Test Case Prioritization in Web Applications}

\author{Maral Azizi}
\affiliation{\institution{University of North Texas}}
\email{maralazizi@my.unt.edu}

\author{Hyunsook Do}
\affiliation{\institution{University of North Texas}}
\email{hyunsook.do@unt.edu}

\begin{abstract}
	The use of relevant metrics of software systems could improve
	various software engineering tasks, but identifying relationships among
	metrics is not simple and can be very time consuming. 
	Recommender systems can help with this decision-making process; many applications
	have utilized these systems to improve the performance of their applications.
	To investigate the potential benefits of recommender systems in regression 
	testing, we implemented an item-based collaborative filtering recommender
	system that uses user interaction data and application change history 
	information to develop a test case prioritization technique.
	To evaluate our approach, we performed an empirical study using three web
	applications with multiple versions and compared four control techniques.
	Our results indicate that our recommender system can help improve the 
	effectiveness of test prioritization.  
\end{abstract}


\maketitle
\thispagestyle{empty}

\section{Introduction}
\label{sec:introduction}

Software systems undergo many changes during their lifetime, and often
such changes can adversely affect the software system. 
To avoid undesirable changes or unexpected bugs, software engineers need to  
test the overall functionality of the system before deploying a new release 
of the product. One of the common ways to evaluate system quality 
in a sequence of releases is regression testing.
In regression testing, software engineers validate the software system
to ensure that new changes have not introduced new faults.
However, modern software systems 
evolve frequently, and their size and complexity grow quickly, and thus  
the cost of regression testing can become too expensive~\cite{jeff16}. 
To reduce regression testing cost, many regression testing and maintenance
approaches including test selection and test prioritization~\cite{marksurvey} 
have been proposed.

To date, most regression testing techniques have utilized various 
software metrics that are available from software repositories, such as
the size and complexity of the application, code coverage, fault history 
information, and dependency relations among components. 
Further, various empirical studies have shown that the use of a particular 
metric or combination of multiple metrics can improve the effectiveness of 
regression testing techniques better than others. 
For example, Anderson et al.~\cite{jeff16} introduced a new technique that 
identifies distinct usage patterns of software through telemetry data and 
showed that their technique can reduce regression test execution time 
by over 30 percent compared to traditional prioritization techniques.
However, we believe that, rather than simply picking one metric over another, 
adopting a recommender system, that identifies more relevant metrics by considering 
software characteristics and the software testing environment might provide a better solution.

Recommender systems have been utilized to alleviate the decision making 
effort by providing a list of relevant items to users based on a 
user's preference or item attributes.
For example, companies that produce daily-life applications, such as Netflix, 
Amazon, and many social networking applications~\cite{recomsurvey05},
are adopting recommender systems to provide more personalized services so that
they can attract more users. 
Recently, recommender systems have been used in software engineering areas 
to improve various software engineering tasks. 
For example, Anvik et al. conducted research that applied machine learning techniques
to developers as well as bug history to make suggestions about 
``who should fix this bug?''~\cite{who}.
While many software engineering techniques have started to incorporate 
recommendation systems, no researchers have investigated the use 
of recommender systems in the area of regression testing.

Therefore, we have investigated whether the use of recommender systems can
improve regression testing techniques, in particular focusing 
on test case prioritization. 
To implement the recommender system, we used user interaction
data and application change history information. 
Previous studies have shown that change history information is an effective indicator 
for bug prediction~\cite{raimund, method}, and the most frequently accessed components 
have a higher impact on the user-perceived reliability of the application~\cite{bryce08, jeff16}. 
Using this kind of information, our recommender system identifies potential components 
that contain faults that can lead to system failure.
By running test cases that exercise such components earlier, 
we can in turn find  defects that are exposable by user interactions earlier.
We implemented a test case prioritization technique by applying our recommender system
and performed an empirical study using two open source and one commercial web applications.
The results of our study show that our proposed recommender system approach can 
improve the effectiveness of test case prioritization 
compared to four other control techniques.

\begin{figure*}[!hb]
\vspace*{-4pt}
        \centering
        \includegraphics[width=0.85\linewidth]{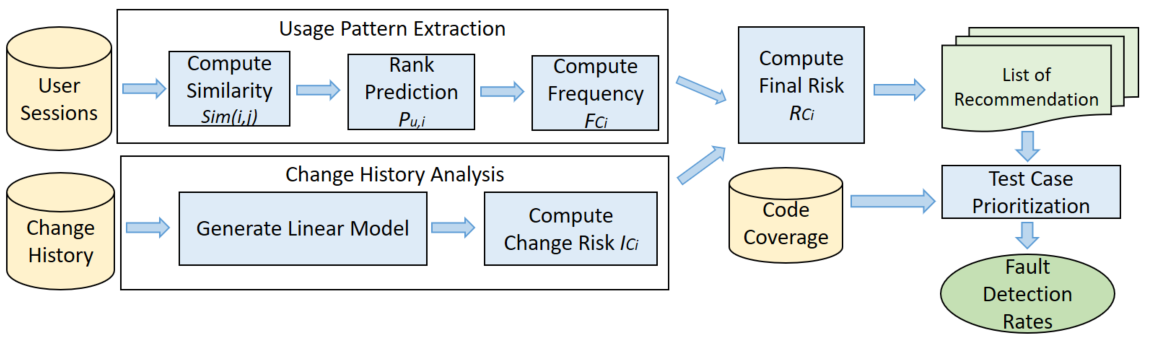}
        \vspace*{3pt}
        \caption{An Overview of Our Test Case Recommender System}
        \label{fig:workflow}
\end{figure*}


The rest of the paper is organized as follows. In Section ~\ref{sec:approach}, we
discuss the approach used in this research and formally define
collaborative filtering recommender systems. 
Sections~\ref{sec:study} and \ref{sec:data} present our empirical study,
including the design, results, and analysis.
Section~\ref{sec:discussion} discusses the results and the implications of these results. 
Section~\ref{sec:related-work} presents background and related work, and
finally in Section~\ref{sec:conclusions}, 
we provide conclusions and discuss future work.

\section{The Proposed Approach}
\label{sec:approach}

Figure~\ref{fig:workflow} shows three major activities in our approach and
how these activities are related to each other. 
The first step of our proposed technique is usage pattern extraction, 
which is shown in the upper box of Figure~\ref{fig:workflow}. 
In this activity, we analyze the users' interaction data to 
determine the most frequently accessed components of the system. 
Our second activity is change history analysis, which is shown in the
lower box of Figure~\ref{fig:workflow}. In this activity,  
we build a classification model to 
measure the relationship between software defects and change history metrics. 
Then, by obtaining the output from these two activities, we measure the risk 
score of each component and prioritize the test cases based on their coverage
of these risky components. In this paper, component refers to method. 

\begin{figure*}[!hb]
\vspace*{-12pt}
	\centering
	\includegraphics[width=0.8\linewidth]{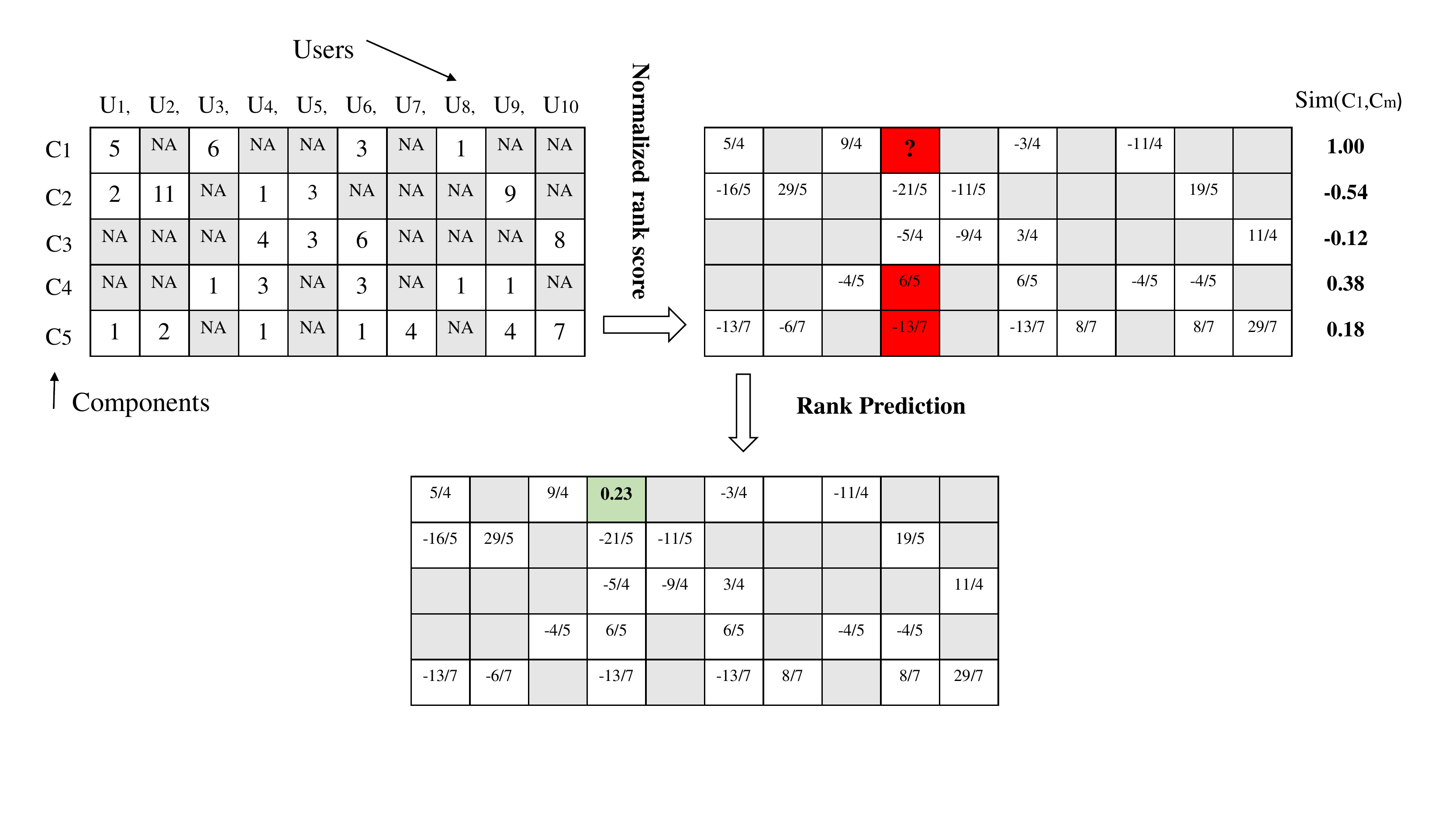}
\vspace*{-30pt}
	\caption{Item-Based Collaborative Filtering Process for Identifying Most-Frequent Components}
	\label{fig:collaborataivefiltering}
\end{figure*} 

\subsection{Usage Pattern Extraction}
\label{recommender-approach}

The goal of our recommender system is to suggest the highest-risk 
components with the most access frequency among all other components in the applications. 
In large scale applications, there are wide ranges of features and components; 
however, in reality, only a relatively small subset of components is accessed by users. 
Therefore, even if there are bugs in some part of the system that is not generally accessed by  users, we assume that those bugs have less impact on the user-reliability perception
of a system; catching those bugs that have been exposed by users is a bigger priority.
 

To calculate the access frequency of the components,
we used two collected data sets: 
a list of users $U=\{u_1, u_2, ..., u_m\}$ and 
a list of the components $C =\{c_1, c_2, ..., c_n\}$. 
Each component $c_i$  has a list of users' ratings if 
users have performed at least one task with it.
Typically, in recommender systems, prediction is based on the numerical values of 
ratings from active users, but in our case we do not have access to such rating 
modules; instead, we use the value of access frequency for a specific component 
by an individual active user as a rating score. 



Suppose that we have a web application that has several functionalities; 
a group of users shows similar interests in using a set of components,
while other groups of users use different sets of components. 
We want to measure the similarity between components by considering users' activities 
and their preferences in using the components.
 
Figure~\ref{fig:collaborataivefiltering} illustrates an example of
component similarity prediction. The upper left-hand matrix in the figure shows 
how many times each component has been used by users. 
The numbers in the cells show the access frequency by the
user $u_{i}$ of the component $c_{j}$, 
and $NA$ indicates that the user $u_{i}$ has not used that 
particular component yet. In this figure, $u_{1}$, $u_{3}$, $u_{6}$, and $u_{8}$ 
utilized components $c_{1}$; and $u_{1}$, $u_{2}$, $u_{4}$, $u_{5}$, and $u_{9}$ 
utilized $c_{2}$. 
We refer to a set of component ratings as a component vector.
For example, the vector $c_1$ is $\langle$ 5, NA, 6, NA, NA, 3, NA, 1, NA, NA$\rangle$
Once we identify the component vectors, we measure the similarity 
between $c_{i}$ and all other components one by one. 

In order to determine the similarity between two components $i$ and $j$, we use
Pearson-r correlation~\cite{recom}.
If $U$ is the set of users who rated components $i$ and $j$, then we compute
the correlation similarity using the following equation:

\vspace*{-3pt}
\[
{Sim (i,j) = \frac {\sum_{u\in {U}}(C_{u,i} - \bar{C_i})(C_{u,j} - \bar{C_j)}}
	{ \sqrt{\sum_{u\in {U}}({C_{u,i} - \bar{C_i}})^{2}}
		{ \sqrt{\sum_{u\in {U}}({C_{u,j} - \bar{C_j}})^{2}}}}}	
\]

where $C_{u,i}$ is the value of access frequency for component $i$ by user $u$, and
$\bar{C_i}$ is the average access frequency value of $C_i$ . 

The reason for finding the similarity between components is to find the missing 
value in the matrix. To measure the similarity between components, 
we normalize the rating of each component by subtracting the row mean. 
The upper right-hand matrix in Figure~\ref{fig:collaborataivefiltering} shows the normalized
rating values of the left matrix. 
For example, the average of access frequency of $c_{1}$ and $c_{2}$  is $15/4$ and $26/5$, respectively.
Then, we subtracted mean values from each corresponding row of the left matrix.  
Positive values indicate that the user used the component more than average;
while negative values
indicate that the user used the component less than average; 0 indicates the average access 
frequency for a component. 
We treat blank values as 0.
The rightmost column in the upper right-hand matrix shows the similarity 
between $c_{1}$ and all other
components. For example, $Sim(c_{1},c_{2})= -0.54 < Sim(c_{1},c_{4})= 0.38$ means
that the probability of rating of $c_{1}$ is much more like $c_{4}$ than $c_{2}$.  

After calculating the similarity between all components, we select 
the set of $N$ most similar components for $c_{i}$;  
this process will be iterated for every component. 
Once we have this set of $N$, then we can make a prediction of access frequency 
for the missing values of $c_{i}$ based on a rating of the $N$ similar components.
To estimate the access frequency rates of ignored components, we performed
ratio prediction computation using a weighted sum method.
This method provides the ratio prediction of a specific component $i$
for user $u$ based on similar components. 

\vspace*{-3pt}
\[
{P_{u,i} = \frac {\sum_{{all\: similar\: components , N}}(S_{i,N} * R_{u,N})}
	{\sum_{{all\: similar\: components, N}}({S_{i,N}})}}
\]

$R_{u,N}$ is the rating score for user $u$ and $N$ most similar components, 
and $S_{i,N}$ is the similarity score of component $i$ and $N$.
We illustrate this using an example in Figure~\ref{fig:collaborataivefiltering}.
The lower matrix in the figure shows the predicted frequency access for $P_{1,4}$, 
which is calculated by $(0.38 * 6/5 + 0.18 * -13/7) / (0.38 + 0.18) = 0.23$.   

The number, $N$, is determined based on the context of application domain, 
ranging from 1 to size of $C - 1$. 
However, assigning a large number to 
$N$ will increase the calculation cost significantly, while the result accuracy would not
change noticeably. Therefore, to reduce the cost of the calculation process, 
we set $N=2$ , 
which means that to predict the missing values we only
select the two most similar components to $c_i$. 
This process is repeated until we find the ranking for the all missing values. 
Once we calculate ratio scores for the components, we can produce a matrix of components
and their access frequency ranking scores.  
We calculate normalized access frequency scores
using the following equation:

\[
{F_{Ci} = \frac {{\sum}(P_{Ci})}
	{{number\: of\: components}}}
\]

where $P_{Ci}$ is the predicted rank score and $F_{Ci}$ is the
normalized rank score of component $i$. 
Then we can sort the matrix of components based on their ranking to 
select $Top-N$  most frequently accessed components.

\vspace*{-3pt}
\subsection{Change History Analysis}
\label{CIA-approach}
\vspace*{-2pt}

The second phase of our proposed approach is change history analysis.
Among hundreds of attributes of code and change history metrics to evaluate
the code quality and error proneness, we chose change history metrics to identify
the riskiest components. According to a previous study~\cite{raimund},
change history is a better indicator for bug prediction purposes than code metrics.

The process of change analysis involves two major steps. 
First, we collect change history information (e.g., added lines of code,
deleted lines of code, etc.) and bug reports
from all available versions of the
applications from their repositories.  
The details of collecting the change history information 
are discussed in Section~\ref{data-collection}.  
Once the change history data is collected, 
we then design a linear model from a set of collected 
change metrics to build a classification model for bug prediction.
Our goal is to find the correlation coefficient of each 
metric to measure statistical relationships between a change metric and 
real defects. 
The value of this measure 
ranges between 1 and -1, where 1 indicates a strong 
positive relationship, 0 indicates no correlation, and a
negative value indicates a reverse correlation.

To evaluate our linear model, we applied 10-fold cross validation
and repeated this process 100 times. 
We used a common accuracy indicator to determine the accuracy of 
our model. 
The three accuracy indicators that we used are PC, TP and FP. 
PC indicates the percentage of correctly classified 
instances,
TP (true positive) indicates the number of components
that contain a bug (and our classification
model also classified them as buggy components), 
and FP (false positive) is the number of components that 
are classified as buggy (but they are clean). 
The output of our classification model is a list of components with
their risk values, $I_{Ci}$ (the risk score of being buggy for component $i$).

	

\vspace*{-3pt}
\subsection{Test Prioritization Using the Recommender System}
\label{test-prioritization}

After obtaining the two metrics explained in previous subsections (component risk scores and 
access frequency ratios), we calculate the final risk scores using the following equation:

\vspace*{-3pt}
\[
{ R_{Ci} = F_{Ci} * I_{Ci}}	
\]

\vspace*{-3pt}
where $F_{Ci}$ is the access frequency score of component $i$ 
and $I_{Ci}$ is the fault risk score of $C_i$.  
Using $R_{Ci}$ scores, our recommender system provides a ranked list of components.
The ranked list of components contains those components of the system that are most
likely to be the cause of regression faults. 
As shown in Figure~\ref{fig:workflow}, the test case prioritization algorithm
reads two inputs (a recommended $Top-N$ list of components and code coverage of tests),
and reorders test cases. 

For example, suppose we have five components
$C =\{c_1, c_2, ... , c_5\}$ with risk scores 
$R =\{0.0014, 0.251, 0.034 , 0.561, 0.138\}$.
Also, suppose we have a list of test cases with their code coverage information
, such as $TC =\{T_1 ={(c_5, c_3)}, T_2={c_1}, T_3={(c_4, c_2)}, T_4={(c_2,c_5)} , T_5={c_2}\}$ .
Then, we reorder the test cases in this order:
$T_3, T_4, T_1, T_2, T_5$, 
since $T_3$ covers $c_4, c_2$, the two components 
having the highest risk score (0.561 + 0.251 = 0.812), and $T_2$ and $T_5$ will be
executed last because $T_2$ covers $c_1$, which has the lowest risk score (0.0014).
$T_5$ only covers $c_2$ (this component has been covered by $T_3$ as well); and because 
$T_4$  covers more components (0.251 + 0.138 = 0.389), then it has higher priority than 
$T_5$, which covers only $c_2 = 0.251$.
Finally, we calculate the fault detection rate of the reordered test cases by 
applying the proposed technique in the latest version of each application.

\section{Study}
\label{sec:study}

Our study investigates whether the use of a recommender system 
can improve the effectiveness of test case prioritization techniques
considering the following research question.

\begin{smallitem}
\item RQ: Can our recommender system be effective in improving
the effectiveness of test case prioritization techniques 
when we have a limited time budget?
\end{smallitem}


\subsection{Objects of Analysis}
\label{sec:objects}

To investigate our research question, we performed an empirical study 
using two open source applications and one commercial web application.
\textbf{DASCP} is a digital archiving and scanning software designed for civil projects; 
we obtained this application from a private company.  
DASCP is a web based application written in ASP.Net and designed to store civil project 
contracts, which include the technical information of civil and construction projects 
such as project plans and relevant associated information. 
Our second application is \textbf{nopCommerce}, which is a widely used open 
source e-commerce web application with more than 1.8 million 
downloads. This application is written in ASP.Net MVC and uses 
Microsoft SQL Server~\cite{nopCommerece}. 
The last application is \textbf{Coevery}, which is an open source 
customer relationship management (CRM) system written in ASP.Net. 
This application provides an easy framework for users to create their own customized 
modules without having to write any code~\cite{coevery}.

\begin{table}
\caption{Application Properties}
\vspace*{-10pt}
\begin{center}
\begin{tabular}{|l|c|c|c|} \hline
\textbf{Metrics}  & \textbf{DASCP} & \textbf{nopCommerce} 
& \textbf{Coevery} \\\hline \hline
Classes   & 107  & 1,919& 2,258 \\\hline
Files  & 201  & 1,473 & 1,875 \\\hline
Functions & 940  & 21,057 & 13,041 \\\hline
LOC & 35,122 & 226,354 &120,897 \\\hline
Sessions  & 748 & 1310 & 274 \\\hline
Faults  & 35 & 70 & 30\\\hline
Versions  & 3 & 23 & 3 \\\hline
Test Cases & 95& 543 & 1,120 \\\hline
Installations & 3 & 2 & 1 \\\hline
\end{tabular}
\end {center}
\vspace*{-15pt}
\label{tab:AUTs}
\end{table}

Table~\ref{tab:AUTs} lists the applications under study and
their associated data: ``Classes'' (the number of class files), 
``Files'' (the number of files), ``Functions'' (the number of 
functions/methods), ``LOC'' (the number of lines of code), 
``Sessions'' (the number of user sessions that we collected), 
``Faults'' (the total number of seeded and real faults), ``Version'' (the number 
of versions), ``Test Cases'' (the number of test cases), and
``Installations'' (the number of different locations where the 
applications were installed). 
Test cases were in the application package; we did not generate any 
new test cases. We downloaded all available versions of open source applications 
from the applications' \textit{GitHub} repositories. 

\subsection{Variables and Measures}
\label{sec:measures}

\subsubsection{Independent Variable}

To investigate our research question, we manipulated one independent 
variable: prioritization technique. 
We considered five different test case prioritization techniques,
which we classified into two groups: control and heuristic.
Table~\ref{tab:techniques} summarizes these groups and techniques.
The second column shows prioritization techniques for each group. 
As shown in Table~\ref{tab:techniques}, we considered four control techniques and 
one heuristic technique. For our heuristic technique, we used the approach 
explained in Section~\ref{sec:approach},
so, here, we only explain the control techniques we applied as follows:

\begin{table}
	\caption{Test Case Prioritization Techniques}
	\vspace*{-10pt}
	\begin{center}
		\begin{tabular}{|l|l|}\hline
			Group & Technique  \\ \hline
			\multirow{4}{*}{Control} 
			& Change history-based ($T_{ch}$)\\
			& Most frequent methods-based ($T_{mfm}$)\\
			& Random ($T_{r}$)\\	
			& Greedy ($T_{g}$)\\\hline			
			\multirow{1}{*}{Heuristic} 
			& Hybrid collaborative filtering-based ($T_{hcf}$)\\\hline
		\end{tabular}
		\end {center}
		\label{tab:techniques}
		\vspace*{-10pt}
	\end{table}

\begin{enumerate}
	\item Change History-Based ($T_{ch}$):
	In order to perform this technique, we used the information that
	we obtained from the change history analysis approach, which we explained in 
	Section ~\ref{CIA-approach}. We prioritized our test cases based on the 
	highest scores of the change risk list. 	
	\item Most Frequent Methods-Based ($T_{mfm}$):
	The most frequent methods usually have high dependency on other classes and methods. 
	If one of them fails, it can cause a significant failure or degradation of the system. 
	In order to prevent a domino effect in the system, high-frequency methods 
	should be tested first, because their failure can cause 
	other components to fail due to their dependencies.	 	
	\item Random ($T_{r}$):
	The random prioritization technique randomly reorders test cases.	
	\item Greedy ($T_{g}$):
	The greedy technique reorders test cases based on the total number of 
	methods covered by test cases. 
	
\end{enumerate}

%
%
%

\subsubsection{Dependent Variable} 

Our research question seeks to measure the effectiveness of our proposed approach
when we have constrained resources.
Qu et al.~\cite{myra} defined the normalized metric of APFD ~\cite{apfd}, which is the
area under the curve when the numbers of test cases or faults are not consistent. 
The NAPFD formula is as follows:

\vspace*{-5pt}
\[
{NAPFD = p- \frac {{TF_{1} + TF_{2} + ... + TF_{m}}} {nm} + \frac{p}{2n}}
\]

In this formula, $n$ is a percentage of the test suite run, 
$m$ represents the total number of faults found by all test cases,
$TF_{i}$ is the first test case that catches the fault $i$,
and $p$ is the number of faults detected by the percentage of our
budget divided by the total number of detected faults when 
running 100\% of test cases.

\subsection{Data Collection and Experimental Setup}
\label{data-collection}
In order to perform our experiment, for both the control and heuristic techniques
we needed to collect three different types of datasets: telemetry data, change 
history, and code coverage information. We explain the data collection processes
in the following subsections.

\subsubsection{Collection of Telemetry Data}
To collect telemetry data, we implemented a small function to record user interactions. 
We considered a sequence of each user's interactions on a specific date as a user session.
First, we uploaded two applications, {\em Coevery} and {\em nopCommerce}, on an IIS server 
at the University of X in November 2016. 
The server specification is CPU Core i7, with 16 GB of RAM.
After deploying our applications, we recruited volunteer graduate and undergraduate
computer science students and assigned a variety of tasks to them. 
These tasks to the volunteers were simple scenarios that each application is designed for.
For example, in {\em nopCommerce}, we asked the volunteers to perform online shopping 
following the actual necessary steps, from login to payment. We also asked some of 
the users to be the system administrator, so we could monitor the whole system 
rather than only the end user side.
We asked the end users to access other parts of the system randomly, for example,
checking their inbox or wish list.  
In total, seventy volunteer students performed different tasks during a forty-day period.  
We collected 1,310 and 274 user sessions for {\em nopCommerce}, 
and {\em Coevery} respectively. 
 
The data collection process for {\em DASCP} is different from that of {\em Coevery} 
and {\em nopCommerce}. 
The {\em DASCP} users whose data we examined are real users, and they have application domain knowledge.
We collected a twelve-month period of user interactions for {\em DASCP}.
In total, 748 user sessions were collected during the designated period of time. 

However, the length of the sessions varied by user, date, and workplace.
For example, some users performed all their assigned task a few hours before the
determined deadline, while others distributed their tasks over several days.
The average length of user sessions for {\em nopCommerce} is equal to 56 and, for
{\em Coevery}, 24. However, in some cases we obtained session lengths over 
300, especially when the interaction date was close to the deadline. 
Also, most of the {\em Coevery} users were selected among graduate students, since the
functionality of this application is relatively more complex that of {\em nopCommerce}. 

\begin{figure}[!ht]
\vspace*{-3pt}
	\centering
	\includegraphics[width=0.90\linewidth]{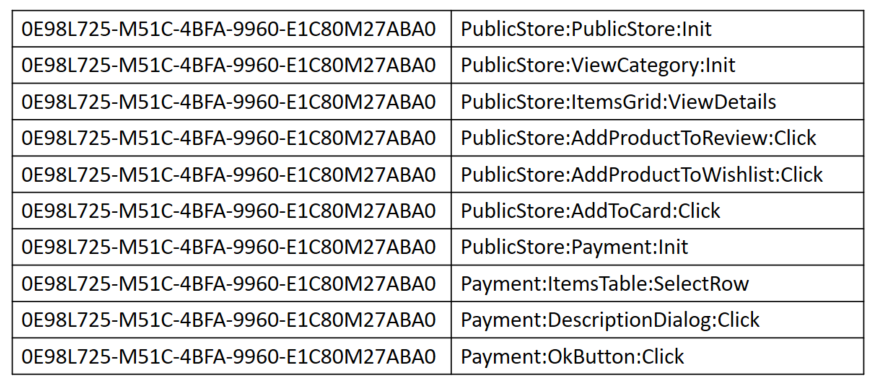}
	\caption{Sample User Session}
	\label{fig:SampleSession3}
\end{figure}

Figure~\ref{fig:SampleSession3} shows an example of the raw telemetry data. 
The left-hand column shows the session identifier, which is a user
navigating through the system. The right-hand column is the set of server-side user 
interactions. The structure of the interactions is of the format (Form name):(Control name):(Action name).

\subsubsection{Collection of Code Change History}
\begin{figure}[!ht]
	\centering
	\includegraphics[width=0.95\linewidth]{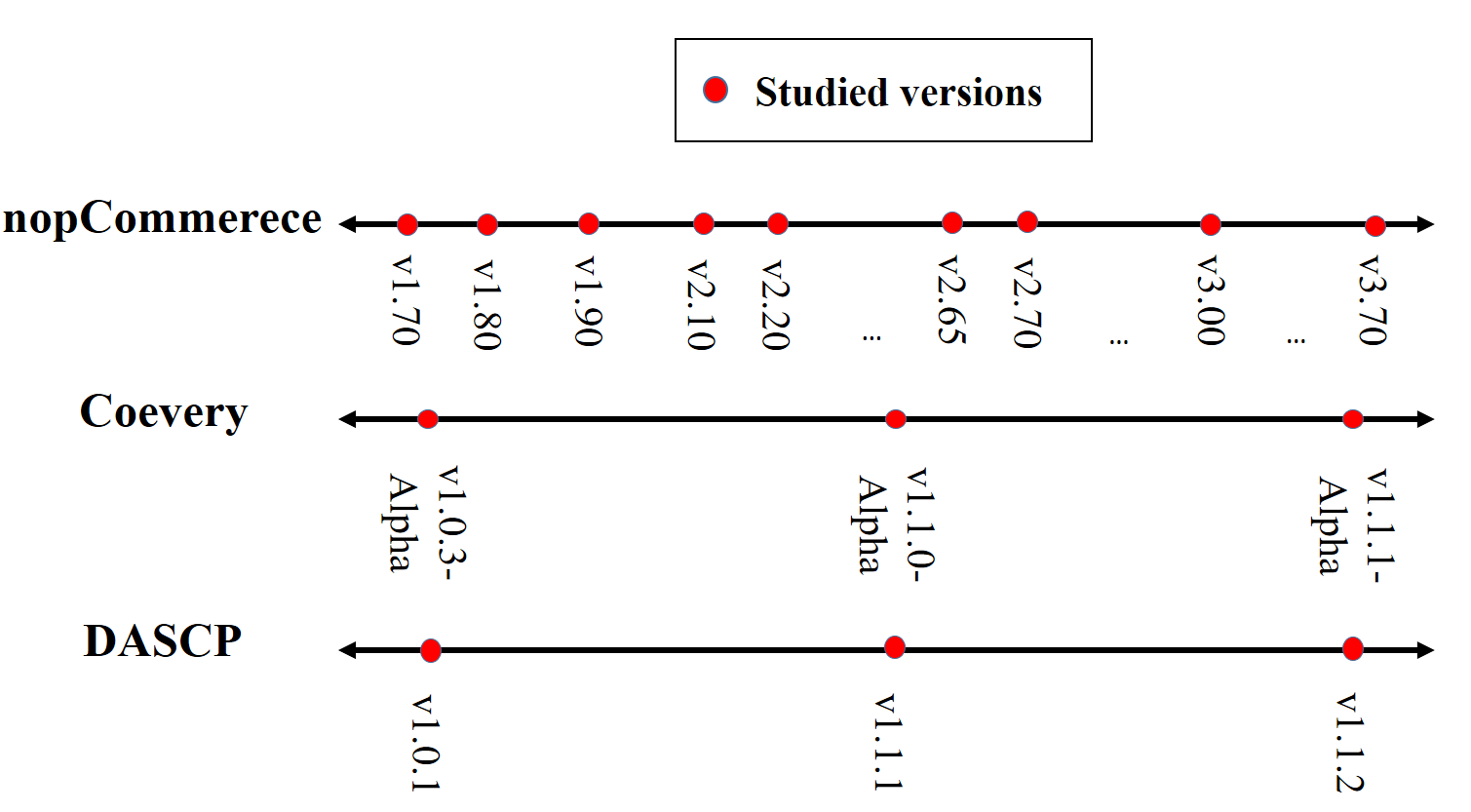}
	\vspace*{-7pt}
	\caption{Versions of Each Application with Change Information and Bug Reports}
	\label{fig:versions}
\end{figure}

We had to take three steps to measure change risk. 
First, we needed a clear understanding of the applications with respect to their changes.
For instance, we needed to check whether a change 
was just the renaming a variable or component, the addition of some comments, 
or an alteration of code by adding or deleting functions, and so on. 
Then, we needed to check whether changes had been made in the current version, 
and finally, we tested a recently changed system~\cite{change3}. 
Figure~\ref{fig:versions} shows the versions that we used in this study. 
{\em nopCommerece} contains 36 versions but we used only the
versions available on the ~\textit{GitHub} repository as of the experiment date. However, for
{\em Coevery} and {\em DASCP}, we used all available versions. 
All studied applications contain fine-grained changes, and the commits on these two open source
applications are available in the \textit{GitHub} repository.   

\textbf{}

In our study, we collected the change history of our three applications 
following Giger et al.'s approach ~\cite{method}. 
Most of these metrics have been used in bug detection research, and they
are known to be good indicators for locating bugs~\cite{sungmicro, shihab12, 
raimund, change1, change2, method}.
Table~\ref{tab:historyMetrics} shows the applied change metrics in this study. 

\begin{table}[!ht]
\caption{Change Metrics Used to Evaluate Risk}	
\vspace*{-3pt}
\begin{tabular} {|l|l|} \hline
	\textbf{Metrics Name} & \textbf{Description} \\\hline \hline
	Modification & Number of times a method was changed  \\\hline
	\multirow{2}{*}{LOC Added} & \parbox[t]{5.5cm}{Added lines of code to a  method \\ body over all method histories} \\\hline
	\multirow{2}{*}{Max LOC Added} & \parbox[t]{5.5cm}{Maximum added lines of code to a method \\ body for all method histories}\\\hline
	\multirow{2}{*}{AVE LOC Added}  & \parbox[t]{5.5cm}{Average added lines of code to a method \\ body per method history}\\\hline
	\multirow{2}{*}{LOC Deleted}  & \parbox[t]{5cm}{Deleted lines of code from a method \\ body over all method histories} \\\hline
	\multirow{2}{*}{Max LOC Deleted} & \parbox[t]{5cm}{Maximum deleted lines from a method \\ body for all method histories}\\\hline
	\multirow{2}{*}{AVE LOC Deleted} & \parbox[t]{5cm}{Average deleted lines from a method \\ body per method history}\\\hline
	Code Churn & Sum of all changes over all method histories\\\hline
	Max Code Churn & Maximum code churn for all method histories \\\hline
	AVE Code Churn & Average code churn  per method history\\\hline
	Age & \parbox[t]{5cm}{Age of a method in weeks from last release} \\\hline
\end{tabular}
\label{tab:historyMetrics}
\vspace*{-6pt}
\end{table}

\subsubsection{Collect Code Coverage}
\label{codecoverage}
Once our recommender system was designed and implemented, 
we needed to find test cases that covered the recommended components. 
We collected the code coverage data of the latest version of AUTs for our test cases using the code coverage analysis tool  
that Microsoft Visual Studio provides as part of its framework. 
After collecting the code coverage information, 
we entered it into a relational database. We assigned unique
identifier values for each method and test case, which provides
a key for method and test tables. Hence, we could easily map the methods
to the test cases that exercise them.

\begin{table}[!ht]
\caption{Code Coverage Data Table}
\vspace*{-10pt}
\begin{center}
\begin{tabular}{|c|c|c|c|c|c|} \hline
	MethodID  & Risk Score & TestID1 & ... & TestID n \\\hline
	12 & 0.876 & 0 &  & 0 \\\hline
	287 & 0.012 & 1 &  & 0 \\\hline
	301 & 0.547 & 0 &  & 0 \\\hline
	148 & 0.145 & 1 &  & 1 \\\hline
	67 & 0.055 & 1 &  & 0 \\\hline			
\end{tabular}
\end {center}
\label{tab:coverage}
\vspace*{-10pt}
\end{table}

Table~\ref{tab:coverage} show the code coverage data we collected.
The first column, ``MethodID'', shows
the unique identifier values that we assigned for each method. 
The second column shows the final risk scores, which is the output of 
our recommender system. Other columns list our test cases
with Boolean values: 0 indicates that the test case does not cover 
the method, and 1 indicates that the test case covers the method.

\subsubsection{Faults}
\label{faultsInfo}

The applications contain real defects reported by users, but the number of 
defects is rather small considering the size of the applications we studied.
Thus, two graduate students seeded additional faults by simulating  programmers' 
common mistakes (e.g., using a single equal sign to check equality).
All seeded faults were in the server-side code level, and we ignored HTML-based and GUI faults. 
Four types of faults were seeded: (1) data faults that are related to interaction with 
the data store; (2) logic faults that are logic errors in the code; (3) action faults that modify 
parameter values and actions; and (4) linkage faults that change the hyperlinks references.  
Table~\ref{tab:AUTs} shows the total number of faults that including both real and 
seeded faults. The number of real faults for {\em nopCommerce}, {\em Coevery}, and 
{\em DASCP} is 25, 8, and 13, respectively. Once we collected all the required data, 
we ran control and heuristic 
techniques and calculated all the dependent values for the reordered test cases.

\section{Data and Analysis}
\label{sec:data}

\begin{table}
	\caption{NAPFD Scores on Average.}
	\vspace*{-10pt}
	\begin{center}
		{\scriptsize
			\begin{tabular}{|c|c|c|c|c|c|c|}\hline
				Application & Test Exe. & \multicolumn{5}{c|} {Techniques} 
				 \\\hline \hline
				& Rate (\%)  & $T_{ch}$ & $T_{mfm}$ & $T_{r}$ & $T_{g}$ & $T_{hcf}$ 
				\\\hline \hline
								
				&10	&18.54	&15.12	&14.33	&19.81	&23.37	\\
				&20	&21.31	&16.78	&15.51	&24.62	&29.98	\\
				&30	&28.86	&25.12	&18.47	&38.23	&40.95	\\
				&40	&40.42	&38.68	&25.59	&43.76	&55.3	\\
				DASCP &50	&54.34	&42.9	&49.53	&50.13	&67.07	\\
				&60	&61.7	&50.34	&45.29	&65.8	&70.42	\\
				&70	&68.34	&65.97	&60.74	&74.66	&76.67	\\
				&80	&75.41	&71.14	&64.88	&79.26	&84.01	\\
				&90	&83.35	&77.69	&67.11	&88.32	&89.83	\\
				&100 &90.16	&79.22	&70.91	&92.39	&94.14	\\\hline \hline
				
				&10	&17.54	&15.75	&8.33	&16.85	&28.14	\\
				&20	&29.28	&28.35	&12.51	&32.39	&47.9	\\
				&30	&38.86	&35.45	&15.57	&34.73	&55.28	\\
				&40	&44.42	&48.68	&27.59	&39.96	&65.06	\\
				nopCommerce &50	&58.34	&54.94	&35.91	&47.65	&74.78	\\
				&60	&62.7	&57.08	&41.45	&53.22	&81.49	\\
				&70	&68.14	&59.97	&50.02	&59.39	&86.38	\\
				&80	&76.22	&64.14	&58.15	&67.41	&89.87	\\
				&90	&79.4	&70.22	&60.17	&71.81	&95.14	\\
				&100 &82.16	&76.04	&63.91	&73.72	&97.06	\\\hline \hline
				
				&10	&26.54	&29.12	&20.33	&26.15	&41.02	\\
				&20	&44.28	&45.12	&24.51	&46.82	&66.98	\\
				&30	&60.86	&50.12	&28.57	&53.71	&73.28	\\
				&40	&63.42	&53.68	&31.59	&65.33	&75.06	\\
				Coevery &50	&64.34	&58.3	&39.62	&67.14	&77.1	\\
				&60	&66.7	&60.41	&44.09	&70.23	&78.65  \\	
				&70	&68.19	&62.97	&46.11	&73.14	&81.13	\\
				&80	&70.67	&65.14	&57.91	&74.09	&83.2	\\
				&90	&71.81	&66.03	&59.01	&74.69	&86.69	\\\hline

			\end{tabular}
		}
		\end {center}
		\label{tab:napfd}
		\vspace*{-5pt}
	\end{table}

Our research question investigates whether the use of the recommender system can improve the 
effectiveness of test prioritization when we have a limited budget for testing,
a common situation that the software industry often faces.
To analysis this research question, we measured the NAPFD, which is a normalized ratio of APFD, when 
our resources were not consistent.
In this experiment, first we executed 10\% of our test cases, and we continued to execute the
test cases in increments of 10\% of the total until they had all been  
executed to see whether we could improve the fault detection rate, given a 
time constraint dictating that running 100\% of the test cases at one time was not feasible. 

Table~\ref{tab:napfd} shows the results of our three applications.
By examining the numbers in the table, we can observe that the improvement
rates of our heuristic technique over the control techniques vary widely. 
When we compared the heuristic with $T_{ch}$, the improvement rates ranged from
4\% to 41\% for {\em DASCP}, from 17\% to 63\% for {\em nopCommerce}, and from 17\% to 54\%
for {\em Coevery}.
When compared with $T_{mfm}$, the improvement rates ranged from
15\% to 78\% for {\em DASCP}, from 27\% to 78\% for {\em nopCommerce}, and from 27\% to 48\%
for {\em Coevery}, indicating results similar to those for as $T_{ch}$.
When compared with $T_{g}$, the improvement rates ranged from
1\% to 33\% for {\em DASCP}, from 31\% to 67\% for {\em nopCommerce}, and from 10\% to 56\%
for {\em Coevery}, showing improvement over a popular and commonly used technique.
As for the comparison with $T_{r}$, the results were more remarkable.
The rates ranged from 32\% to 282\% for all three applications. 
Note that the results for random technique are average of iterating 
this technique for 10 times.

One outstanding trend we observed in the table is that the improvement 
rates are much higher when the time budget is smaller.
For example, in the comparison with $T_{ch}$ for {\em nopCommerce}, 
when 10\% of the budget was assigned, the improvement rate was 60\%,
but when we had a full budget, the rate dropped to 18\%.
A similar trend can be observed across all control techniques and applications.  
This indicates that our approach can be more helpful when companies are operating
under a tight budget.

\section{Discussion}
\label{sec:discussion}



Our experiment results indicate that by utilizing a recommender system
when we prioritize test cases, we were able to improve the effectiveness 
of test case prioritization. 
While we have not conducted a cost benefit analysis of our approach,
which is our future work, we found that the cost of applying our approach is
negligible. The biggest cost for applying our approach is calculating the matrix
of access frequency scores, but it took only 580, 153, and 374 seconds
for {\em nopCommerce}, {\em Coevery} and {\em DASCP}, respectively.  
Further, the prioritization algorithm took only 0.29 seconds for {\em nopCommerce},
the application that has the largest number of test cases.

\vspace*{1pt}
\subsubsection*{Limitations of Applying Recommender Systems}

There are three common limitations in collaborative filtering recommender systems;
new user problem, new item problem, and sparsity~\cite{recomsurvey05}. 
Among these three limitations, two of them are related to our study. 

	
	{\textbf{Sparsity.}}	
	As we discussed earlier, we do not have a rating module in our
	applications, and thus we used access frequencies to each component
	as a user rating. Further, for the open source applications, {\em nopCommerce} 
    and {\em Coevery}, we collected the user interaction data
	from nonprofessional users, so our data could contain noise and 
	redundancy, which can affect the performance of our technique. 
	Moreover, the number of collected ratings is relatively small, compared 
	to the number of expected ratings required to generate an accurate prediction.
	
	Also, the distribution patterns of user ratings can affect the outcome
    of collaborative filtering algorithms. For example, in {\em nopCommerce} case, 
	some components were used by all users,
	such as registration and membership components, while some other
	components were ignored by the majority of users. 
	In order to eliminate this issue, we need to collect more user 
    interaction data by considering a larger number of users and a longer 
    period of time.
	In addition, having actual users would generate more realistic results, because
    their interactions would be based on real business functions and system workflows.
	
	{\textbf{New Item Problem.}}	
	Another limitation of collaborative filtering that is related to
	our study is the ``New Item Problem''. It is a common practice that
	newly developed components are frequently added to a system. 
	However, rankings on collaborative filtering algorithms are 
        based on user access frequencies to the components. Therefore, 
        the newly added components would not appear in the recommender suggestion 
        list until a certain number of users perform some tasks on them. 
	In our study, in addition to using access frequency scores, 
    we also applied change risk scores to recommend the riskiest components.
	Therefore, even if a new component has a high change risk score,
    its frequency score would still be zero, which makes the overall 
    risk score zero. To overcome this limitation, we need
	to use a hybrid and normalized ranking score by assigning 
    a small value to the components.

\section{Threats to Validity}
\label{sec:validity}

The primary threat to validity of this study is the amount of
user session data and the type of users who participated in this
study. For the proprietary application that we used in this study,
we collected user interaction data for a long period time; 
the collected data was created by actual users of 
the application. However, for the two open source applications, 
the period of time over which we collected user interactions was relatively 
short, and the participants were not domain experts or regular users 
of the applications, so their usage patterns had wide variations.
This threat can be addressed by performing additional studies that
monitor user interactions over a longer time period among a wider population,
and by considering industrial applications and different types of 
applications (e.g., mobile applications).
 

Another concern involves the bug reports that we used.
Our classification prediction values for designing linear models 
in the change history analysis were generated from bug history that 
was reported by actual users.
Further, using these bug reports, we measured the coefficient of other 
variables to create our linear model for change history analysis.
Because our bug report data is not comprehensive and contains 
only those bugs accrued 
up to the time that we stopped collecting data, 
and because there might 
be other bugs that have not been reported yet or that might occur 
later, there is a possibility that the bug reports are biased.

\vspace*{2pt}
\section{Related Work}
\label{sec:related-work}


\subsubsection*{Test Case Prioritization}
Due to the appealing benefits of test case prioritization in practice, 
many researchers have proposed various  
techniques. These techniques help engineers discover faults
early in testing, which allows them to begin debugging earlier.
Recent surveys ~\cite{catal13, marksurvey} provide a comprehensive 
understanding of overall trends of the techniques and suggest areas for improvement.
Depending on the types of information available, various test case
prioritization techniques can be utilized, but 
the majority of prioritization techniques have used source code
information to implement the techniques.
For instance, many researchers have utilized code coverage information
to implement prioritization techniques~\cite{elbaum02feb, kim02may,rothermel01oct}. 

Further, some researchers have used software risk information in testing approaches.  
For instance, Frankl and Weyuker ~\cite{weyuker} introduced two risk related measures of 
software testing effectiveness, which are expected detected risk and expected risk reduction
and investigated the effectiveness of these two measures on testing techniques. 
Hettiarachchi et al.~\cite{risk} proposed a new test case
prioritization technique. Their technique uses risk levels of potential defects
to detect risky requirements then it prioritizes test cases by mapping the related 
test cases and these requirements.

More recently, several prioritization techniques 
utilizing other types of information have also been proposed. 
For example, Anderson et al. applied telemetry data to compute fingerprints 
to extract usage patterns and for test prioritization~\cite{jeff16}.
Brooks and Memon performed a study applying telemetry 
data to generate usage pattern profiles~\cite{memongui}.  
Gethers et al. presented a method  that uses textual change of source code
to estimate an impact set ~\cite{kagdichange}. 

\subsubsection*{Recommender Systems}  
Recommender systems are software engineering tools that make 
the decision making process easier by providing a list of relevant items.
Some widely-used applications that provide recommender systems
are Amazon, Facebook, and Netflix. These applications provide suggestions
to target users based on the user or on item characteristic similarities.  

With the fast growth of such information, machine learning technologies  
motivate software engineers to apply recommendation systems in software 
development. Recommender systems in software engineering have been applied  
to improve software quality and to address the challenges of the development process~\cite{rssebook}.  
For instance, Murakami et al. ~\cite{murakami} proposed a technique that 
uses user editing activities  detecting code relevant to existing methods. 
Danylenko and Lowe provided a context-aware recommender system 
to automate a decision-making process for determining the efficiency of 
non-functional requirements ~\cite{contextawar}.

As discussed briefly earlier, many types of information are available 
for implementing test case prioritization techniques.
In this research, we collected over 2,000 user sessions from 
three different web applications and gathered the change history of each application. 
Our research seeks to apply item-based collaborative filtering algorithms 
to generate a recommendation list for test prioritization.
To our knowledge, our recommender system-based prioritization technique is novel 
and has not yet been explored in regression testing.

\vspace*{3pt}
\section{Conclusions and Future Work}
\label{sec:conclusions}

In this research, we proposed a new recommender system to 
improve the effectiveness of test case prioritization. 
Our recommender system uses three datasets (code coverage, change history, and user sessions)
to produce a list of the riskiest components of a system for regression testing.
We applied our recommender system using two open source applications and 
one industrial application to investigate whether our approach 
can be effective compared to four different control techniques.
The results of our study indicate that our recommender system
can improve test case prioritization; also, the performance
of our approach was particularly noteworthy when we had
a limited time budget.

In future research, we plan to investigate other approaches 
to address a sparsity problem of our recommender system by applying an associative
retrieval framework and related spreading activation algorithms
to track user transitive interactions through their previous interactions. 
To address the ``New Item Problem,'' we plan to apply 
knowledge-based techniques such as case-based reasoning. 

\subsection*{Acknowledgments}

This work was supported, in part, by NSF CAREER Award
CCF-1564238 to University of North Texas.

\balance
\bibliographystyle{plain}
\bibliography{paper} 


\end{document}